\documentclass[prb,twocolumn,showpacs,aps,superscriptaddress,floatfix]{revtex4-2}
\usepackage{amsmath}
\usepackage{amssymb}
\usepackage{amsfonts}
\usepackage{bm}
\usepackage[dvips]{graphicx}
\usepackage{color}
\usepackage{ulem}
\usepackage{tabularx}
\usepackage[unicode, pdftex]{hyperref}
\usepackage{verbatim}
\DeclareGraphicsExtensions{.eps,.pdf}
\graphicspath{{pictures/}}
\usepackage{epstopdf}
\begin{document}
	\title{Magnetization control of the nematicity direction and nodal points in a superconducting doped topological insulator}
    \author{D. A. Khokhlov}
    \affiliation{Independent researcher} 
   
    \author{R. S. Akzyanov}
    \affiliation{Dukhov Research Institute of Automatics, Moscow, 127055 Russia}
       \affiliation{Moscow Institute of Physics and Technology, Dolgoprudny,
    Moscow Region, 141700 Russia}
    \affiliation{Institute for Theoretical and Applied Electrodynamics, Russian
    Academy of Sciences, Moscow, 125412 Russia} 
    \author{A. V. Kapranov}
    \affiliation{Dukhov Research Institute of Automatics, Moscow, 127055 Russia}
       \affiliation{Moscow Institute of Physics and Technology, Dolgoprudny,
    Moscow Region, 141700 Russia}

	\begin{abstract}
    We study the effects of magnetization on the properties of the doped topological insulator with nematic superconductivity. We found that the direction of the in-plane magnetization fixes the direction of the nematicity in the system. The chiral state is more favorable than the nematic state for large values of out-of-plane magnetization. Overall, the critical temperature of the nematic state is resilient against magnetization. We explore the spectrum of the system with the pinned direction of the nematic order parameter $\Delta_{y}$ in details. Without magnetization, there is a full gap in the spectrum. At strong enough out-of-plane $m_z$ or orthogonal in-plane $m_x$ magnetization, the spectrum is closed at the nodal points that are split by the magnetization. Flat Majorana surface states connect such split bulk nodal points. Parallel magnetization $m_y$ lifts nodal points and opens a full gap in the spectrum. We discuss relevant experiments and propose experimental verifications of our theory.
    \end{abstract}

	\maketitle
\section{Introduction}
Superconductivity in doped topological insulators of Bi$_2$Se$_3$ family was observed in various experiments~\cite{Hor2010,Kriener2011,Sasaki2011,Kriener2012,Kirzhner2012,Shruti2015,Qiu2015,Liu2015,Matano2016,Pan2016,Yonezawa2016,Asaba2017,Tao2018,Kuntsevich2018,Kurter2018,Neha2019,Kuntsevich2019,Neha2019, Das2020,Kawai2020,Almoalem2020,Bagchi2022,Yokoyama2023}. To achieve a superconducting state, bismuth selenide is doped with Cu, Sr, or Nb~\cite{Yonezawa2018}. Finite Knight shift in nuclear magnetic resonance shows that Cooper's pairs have spin-triplet pairing in such materials~\cite{Matano2016}. Superconductivity in doped topological insulators lowers the rotational symmetry of the system from $C_3$ in the normal state to $C_2$ in the superconducting state. This bright feature appears in observation of the second critical field~\cite{Pan2016,Kuntsevich2018}, magnetic resonance~\cite{Asaba2017}, vortex core form~\cite{Tao2018}, specific heat measurements with applied in-plane magnetic field~\cite{Yonezawa2016}, quasiparticle interference~\cite{Chen2018,Khokhlov2021}.

Symmetry-based theoretical investigation shows that the order parameter from the $E_u$ representation of the $D_{3d}$ crystalline group satisfies both experimental observations: spin-triplet pairing and rotational symmetry breaking~\cite{Fu2010, Fu2014,Venderbos2016}. This order parameter is a two-component vector. Real order parameter $\pmb{\eta}=\eta(\cos\alpha;\sin\alpha)$ keeps time-reversal symmetry, transforms as a coordinate vector $(x,y)$ and is called a nematic order parameter. The direction of this vector $\alpha$ defines the direction of the anisotropy of the system and is often called nematicity. Usually, two directions of the nematicity $\alpha=0$ that corresponds to $\Delta_x$ and $\alpha=\pi/2$ that corresponds to $\Delta_y$ are considered. A reach variety of different phenomena has been predicted theoretically for the topological insulators with nematic superconductivity, such as Majorana surface states~\cite{Hsieh2012,Hao2017,Khokhlov2022}, vestigial order~\cite{Hecker2018,How2023,Hecker2023}, unconventional Abrikosov vortices~\cite{Zyuzin2017,How2020}, spin and spin-mass vortices~\cite{Wu2017,Akzyanov2021sv,Kapranov2023}, chiral Higgs modes~\cite{Uematsu2019}, anomalous Josephson Hall effect~\cite{Akzyanov2022}, partial paramagnetic response to the magnetic field ~\cite{Schmidt2020,Khokhlov2021_2}.

In topological insulators, the Fermi surface is hexagonally deformed due to $C_3$ rotational crystal symmetry. This deformation arisies due to cubic in momentum terms in a Hamiltonian that are called as hexagonal warping~\cite{Fu2009,Liu2010}. Such warping significantly affects the properties of the nematic superconductivity.Fo example, hexagonal warping stabilizes nematic superconductivity as a ground state~\cite{Akzyanov2020,Akzyanov2021}. Also, warping can open a full gap in the spectrum and fix a direction of the nematicity of the system along the crystal axes~\cite{Fu2014,Venderbos2016_2}.

Singlet superconductivity can be destroyed by magnetization since the Zeeman field pulls apart electrons with the different spins in Cooper's pair~\cite{Clogston1962,Maki1964}. In the case of spin-triplet superconductivity, system can stay in the superconducting state even for magnetization $m\gg\Delta$ since in a  triplet Cooper's pairs electrons have the same spin~\cite{Powell2003}. In a theoretical article Ref.~\cite{He2018}, authors investigate s-wave superconductivity in NbSe$_2$. Due to strong spin-orbit coupling, the Clogston limit for Cooper's pair breaking is high, and a magnetic field can close the gap while superconductivity survives. In such a system, a nodal-point superconductor with Majorana fermions can appear. In Ref.~\cite{Wong2013}, authors investigate $p$-wave topological superconductor in the presence of external magnetization. Superconductivity survives in such material even for high values of magnetization. Strong magnetization closes the superconducting gap at nodal points connected by the flat Majorana bands.

Usually, the direction of the anisotropy in the superconducting state of the doped topological insulators is fixed by the some crystal fields that arise due to finite strain in the normal state~\cite{Kuntsevich2019}, but other possibilities for such pinning remain~\cite{Smylie2022}. However, in the recent experiment~\cite{Yokoyama2023}, it was shown that a strong enough magnetic field can change the system's nematicity direction. This work raises of interesting question of the possibility of the control of the nematicity in doped topological insulators by external magnetization.

In this work, we focus on the effects of the strong magnetization with the arbitrary direction in the doped topological insulator with nematic superconductivity. We write down linearized in the order parameter Gor'kov equations for such a system and calculate the critical temperature. For the in-plane magnetization, we found that the highest critical temperature is realized for the nematic order parameter, which direction is orthogonal to the magnetization. In this case, the critical temperature is independent of the value of the magnetization. If magnetization is collinear to the nematicity direction, then critical temperature decreases with the increase of the magnetization. In implies that direction of the magnetization can fix the nematicity direction. Out-of-plane magnetization decreases critical temperature for the nematic state. The chiral state becomes favorable for a high enough value of the out-of-plane magnetization. Overall, the critical temperature of the nematic and chiral states is finite, even for large magnetization values. We study the spectrum of the nematic state with the pinned direction $\alpha = \pi/2$ (which is $\Delta_{y}$) in details. Without hexagonal warping nematic superconducting state has nodal points at the Fermi energy.
In this case, out-of-plane magnetization $m_z$ transforms nodal points to the Fermi surface. With hexagonal warping, the spectrum is fully gapped without magnetization. Finite out-of-plane magnetization opens 12 bulk nodal points that come in pairs. Each pair is split by the magnetization. Orthogonal in-plane magnetization $m_x$ opens 4 bulk nodal points in pairs. In mixed magnetization, in the $Oxz$ plane, we have 4 of 12 nodal points depending on the projection values of the magnetization on $Oz$ and $Ox$ axes. Parallel in-plane magnetization $m_y$ lifts nodal points and opens a full gap in the spectrum. We calculate a tight-binding spectrum and discover that each split pair of nodal points is connected by flat Majorana surface states. We discuss the possible experimental verifications of our work.

\section{Model}\label{suscp_model}
\subsection{Normal phase}
We describe bulk electrons in a doped topological insulator of Bi$_2$Se$_3$ family by low-energy $\mathbf{k\cdot p}$ two-orbital Hamiltonian~\cite{Liu2010}: 
\begin{eqnarray}	
	\label{Eq::H0}
 	\hat{H}_{0}(\mathbf{k})=-\mu+m\sigma_z+v_zk_z\sigma_y+v(k_xs_y-k_ys_x)\sigma_x+\\ 				\nonumber \lambda(k_x^3-3k_xk_y^2)s_z\sigma_x,
\end{eqnarray}
where $\mu$ is the chemical potential, $2m$ is a single-electron gap at zero chemical potential, Fermi velocities $v$ and $v_z$ describe motion in the $(\Gamma K;\Gamma M)$ plane and along $\Gamma Z$ direction correspondingly, $\lambda$ describes hexagonal warping. In the general model, there is one more term with the hexagonal warping $\lambda_2(k_y^3-3k_x^2k_y)\sigma_y$, see Ref.~\cite{Liu2010}. This term has the same spin and orbital structure as the term $v_zk_z\sigma_y$. Thus, adding hexagonal warping $\lambda_2$ simply transforms the plane $k_z=0$ to a manifold $v_zk_z+\lambda_2(k_y^3-3k_x^2k_y)=0$. Such a transformation makes the calculation more complicated but does not bring any new physics for our calculations. Therefore, we put $\lambda_2=0$ in our model. Pauli matrices $s_i$ act in spin space while matrices $\sigma_i$ act in space of Bi and Se orbitals $\mathbf{p}=(P^1,P^2)$, where $i=\{x,y,z\}$, Planck constant $\hbar=1$. The Hamiltonian~(\ref{Eq::H0}) obeys time-reversal symmetry $\hat{\mathcal{T}} \hat{H}_0(\mathbf{k}) \hat{\mathcal{T}}^{-1}=\hat{H}_0(-\mathbf{k})$, where $\hat{\mathcal{T}}=is_y \hat{K}$, $\hat{\mathcal{T}}^2=-1$  is time-reversal operator and $\hat{K}$ provides complex conjugation. Also, this Hamiltonian has inversion symmetry $\hat{P}\hat{H}_0(\mathbf{k}) \hat{P}=\hat{H}_0(-\mathbf{k})$, where $\hat{P}=\sigma_z$, $\hat{P}^2=1$ is the inversion operator~\cite{Liu2010}. Note, hexagonal warping lowers $C_{\infty}$ rotational symmetry down to $C_3$. 
\subsection{Superconducting phase}\label{subsec::SC_phase}
We describe superconductivity in Nambu-II basis, where the wave function is  
\begin{eqnarray}
    \Psi_{\mathbf{k}}=(\phi_{\mathbf{k}}^t,-i\phi_{\mathbf{k}}^{\dagger}s_y)^t,
\end{eqnarray}
with $\phi_{\mathbf{k}}=(\phi_{\uparrow,1,\mathbf{k}},\phi_{\downarrow,1,\mathbf{k}},\phi_{\uparrow,2,\mathbf{k}},\phi_{\downarrow,2,\mathbf{k}})^t$, symbol $t$ means transposition and symbol $^{\dagger}$ means Hermitian conjugation. Operator $\phi_{\uparrow(\downarrow),\sigma,\mathbf{k}}^{(\dagger)}$ annihilates (creates) electron with up (down) spin on the orbital $\sigma=(P^1,P^2)$ with momentum $\mathbf{k}$. Superconducting order parameter from E$_u$ representation of D$_{3d}$ crystalline point group has vector structure with two components $\pmb{\eta}=(\eta_x;\eta_y)$. It has the following matrix structure~\cite{Venderbos2016_2}:
\begin{eqnarray}
\label{Eq::Delta}
\hat{\Delta}=\eta_x\hat{\delta}_x+ \eta_y\hat{\delta}_y,
\end{eqnarray}
where $\hat{\delta}_{x,y}=s_{x,y}\sigma_{y}$. We can write the order parameter as $\eta_x=\eta\sin(\alpha)e^{i\phi_1}$ and $\eta_y=\eta\cos(\alpha)e^{i\phi_2}$, where $\phi=\phi_1-\phi_2$. Order parameter with $\phi=0$ is called nematic, which can have arbitrary orientation $\alpha$. The nematic order parameter has rotational symmetry $C_2$, breaking $C_3$ crystalline symmetry. This order parameter has time-reversal symmetry. When angles $\phi \neq 0$, time-reversal symmetry is broken. Particularly, order parameter $\eta(\frac{1}{\sqrt{2}};\pm\frac{i}{\sqrt{2}})$ is called chiral~\cite{Venderbos2016_2}. We assume that only electrons in the Debye window participate in the superconductivity $-\omega_D<\epsilon_{\mathbf{k}}<\omega_D$, where $\epsilon_{\mathbf{k}}$ is the band's dispersion of the Hamiltonian~(\ref{Eq::H0}).  

The BdG Hamiltonian in Nambu-II basis is~\cite{Fu2010}:
\begin{eqnarray}
\hat{H}_{BdG}(\mathbf{k})=\tau_z\hat{H}_0(\mathbf{k})+\hat{\Delta}\frac{\tau_x+i\tau_y}{2}+\hat{\Delta}^*\frac{\tau_x-i\tau_y}{2}.
\label{Eq::Hbdg}
\end{eqnarray}
Matrices $\tau_i$ act in electron-hole space. The Hamiltonian~(\ref{Eq::Hbdg}) obeys electron-hole symmetry 
\begin{eqnarray}
    \label{Eq::Xi}
	\hat{\Xi}^{-1} \hat{H}_{BdG}(\mathbf{k}) \hat{\Xi}=-\hat{H}_{BdG}(-\mathbf{k}),
\end{eqnarray}
where $\hat{\Xi}=s_y\tau_yK$ and $K$ is complex conjugation. 

In the absence of the hexagonal warping $\lambda=0$, the system degenerates with respect to the nematicity direction $\alpha$. The superconducting gap has two nodal points at the Fermi energy with coordinates $\mathbf{k}\pm\sqrt{\mu^2+\eta^2-m^2}/v\left(-\sin\alpha;\cos\alpha;0\right)$.

In the presence of the hexagonal warping rotational symmetry of the normal phase becomes $C_3$, and the spectrum is sensitive to the mutual orientation of the warping and the order parameter~\cite{Venderbos2016,Khokhlov2021}. Particularly, the order parameter $\Delta_x$ has nodal points in the spectrum, while the system with $\Delta_y$ has the highest full gap $2\eta\sqrt{1-m^2/\mu^2}$ among possible orientations of the nematicity. 
\subsection{Magnetization}
We consider the Zeeman splitting due to finite magnetization. It can appear due to the proximity effect from magnetic substrate~\cite{Menshov2012,Li2015} or due to a finite magnetic field. The magnetization can be written as~\cite{Liu2010}:
\begin{eqnarray}
	\label{Eq::Hm}
	\hat{H}_m=m_{x}s_x+m_{y}s_y+m_{z}s_z,
\end{eqnarray}
where $m_{i}$ is the strength of the magnetization along axis $i$, where $i=x,y,z$. 
   
The magnetization breaks time-reversal symmetry and lifts the twofold Krademer's degeneracy of the $\hat{H}_{BdG}$. The full Hamiltonian $\hat{H}_{BdG}(\mathbf{k})+\hat{H}_m$ obeys chiral symmetry $\hat{\Sigma}^{-1}\left(\hat{H}_{BdG}(\mathbf{k})+\hat{H}_m\right)\hat{\Sigma}=\hat{H}_{BdG}(-\mathbf{k})+\hat{H}_m$, where operator $\hat{\Sigma}=\sigma_z\tau_z$. Thus, each positive eigenvalue $E_{\mathbf{k}}$ is accompanied by the partner $-E_{\mathbf{k}}$~\cite{Chiu2016}.

\section{Critical temperature in presence of magnetization}
We calculate the critical temperature of the nematic superconductor as a function of magnetization. For simplicity, we present results only for systems without hexagonal warping $\lambda =0$—including the warping leads to some insignificant enhancement of the critical temperature. See Ref.~\cite{Akzyanov2021} for the case without magnetization. 

Matsubara Green's function of the normal state electrons is $\hat{G}_{0,e}(\omega,\mathbf{k})=\left(\omega-\hat{H}_0(\mathbf{k})-\hat{H}_m\right)^{-1}$, where fermionic frequency $\omega=\pi T(2n+1)$. For holes in the normal state we get $\hat{G}_{0,h}(-\omega,\mathbf{k})=\left(\omega+\hat{\mathcal{T}}^{-1}(\hat{H}_0(\mathbf{k})+\hat{H}_m)\hat{\mathcal{T}}\right)^{-1}$. Anomalous Green's function in the linear approximation is $\hat{F}^{(1)}_{\alpha}(\omega,\mathbf{k})=\hat{G}_{0,e}(\omega,\mathbf{k})\hat{\delta}_{\alpha}\hat{G}_{0,h}(-\omega,\mathbf{k})$. We introduce $f_{\alpha\beta}(\omega,\mathbf{k})=\textrm{Tr}\left(\hat{\delta}_{\alpha}\hat{F}_{\beta}\omega,\mathbf{k})\right)$, where trace is taken over spin and orbital degrees of freedom. In the vicinity of the critical temperature order parameter is small. Then, the gap equation can be written in the linearized form:
\begin{eqnarray}
\label{Eq::gap_eq}
 \eta_{\alpha}=\frac{g T}{4}\sum_{\omega}\int\frac{d^3\mathbf{k}}{(2\pi)^3}f_{\alpha\beta}(\omega,\mathbf{k})\eta_{\beta},
\end{eqnarray}
where $g$ is the coupling constant, $\alpha,\beta$ corresponds to $x,y$. Integration is performed over the first Brillouin zone, and summation is provided over fermionic Matsubara frequencies $\omega=(2n+1)\pi T$. Thus, we obtain two linear equations on $\eta_x, \eta_y$. Superconductivity appears when the system has a nontrivial solution. Thus, the system~(\ref{Eq::gap_eq}) should be degenerate, and we get a condition on critical temperature $T_c$. The determinant of the system~(\ref{Eq::gap_eq}) is equal to zero. 

\begin{figure}[h]
\center{\includegraphics[width=1\linewidth]{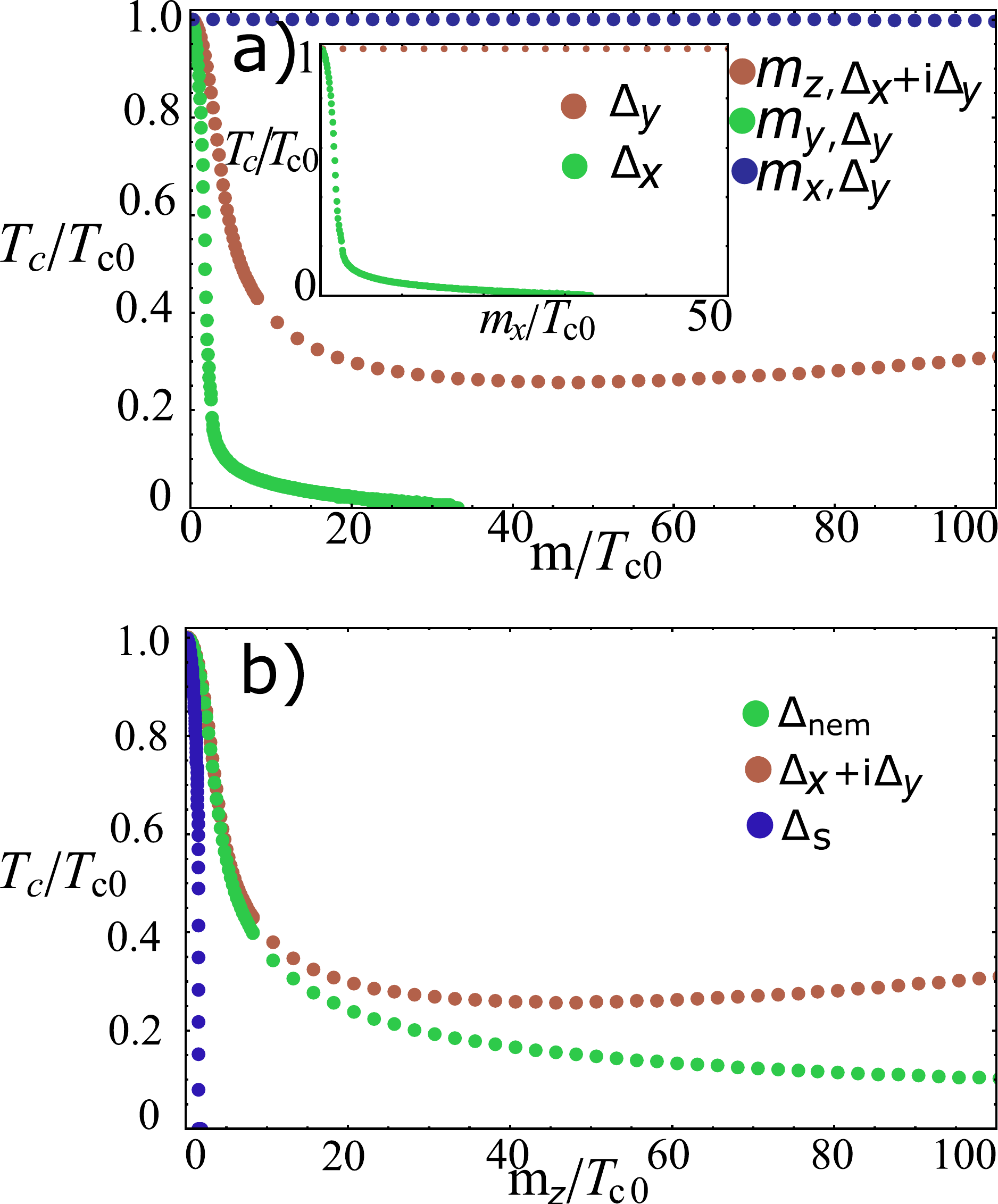}}
\caption{Critical temperature $T_c/T_{c0}$ vs magnetization $m_{z}/T_{c0}$. We set $\mu=2m$ and $\lambda=0$. Panel a): We consider three different orientations of the magnetization. For $m_z$, we plot the critical temperature of chiral phase $\Delta_x+i\Delta_y$. For $m_x$ and $m_y$, we fix the order parameter as $\Delta_y$. Insert shows the critical temperature of $\Delta_x$ and $\Delta_y$ as a function of the $m_x$. Panel b): Critical temperature for s-wave, nematic, and chiral $\Delta_x+i\Delta_y$ phases as a function of the magnetization $m_z$.}
\label{Fig::Tc_vs_M}
\end{figure}

We start with magnetization along the $Oz$ axis. For this geometry critical temperature is defined by $(1-\Phi_d)^2=\Phi_{od}^2$, where $\Phi_{xx}=\Phi_{yy}=\Phi_d=\frac{g T}{4}\sum_{\omega}\int \frac{d^3\mathbf{k}}{(2\pi)^3}f_{\alpha\alpha}(\omega,\mathbf{k})$ and $i\Phi_{od}=\frac{g T}{4}\sum_{\omega}\int \frac{d^3\mathbf{k}}{(2\pi)^3}f_{\alpha\overline{\alpha}}(\omega,\mathbf{k})$. The exact solution is chiral $\Delta_x\pm i\Delta_y$ for any nonzero $\Phi_{od}$ induced by the field $\pm m_z$, where $m_z\neq0$. Further, we consider positive magnetization $m_z>0$. We plot the critical temperature of the chiral phase in Fig.~\ref{Fig::Tc_vs_M} by red dots in panels a) and b). 

Also, we consider the case when nematic order $\Delta_x$ or $\Delta_y$ is pinned. In this case, the critical temperature is determined by $\Phi_d=1$. We calculate the critical temperature for pinned nematic order parameter and show it in Fig.~\ref{Fig::Tc_vs_M}b). The difference between critical temperature for nematic and chiral order parameters is low for $m_z\lesssim 10$ - $20$ $T_{c0}$. The difference becomes larger at higher magnetization. 

The critical temperature slowly decreases towards zero for both chiral and nematic phases as a function of $m_z$. This fundamentally differs from singlet pairing~\cite{Clogston1962,Maki1964}, where a field of several $T_c$ destroys the superconductivity. We calculate $T_c$ for our system and plot it by blue dots in Fig.~\ref{Fig::Tc_vs_M}b). If we calculate the critical temperature for the topological insulator with the s-wave order parameter, we get that critical temperature vanishes if the Zeeman field is of magnitude $\sim1.6 T_{c0}$. Note, s-wave order parameter is independent of the direction of the magnetization. 

Further, we consider the Zeeman field in $(\Gamma M;\Gamma K)$ plane. We focus on the two orientations along $Ox$ or $Oy$. Note, we consider a system without hexagonal warping; thus, $C_{\infty}$ rotational symmetry of normal state presents. In this case parameter $\Phi_{od}=0$ and $\Phi_{xx}\neq\Phi_{yy}$. Such a situation allows $\Delta_{x}$ and $\Delta_{y}$ order parameters. We calculate the critical temperature for $\Delta_{y}$ order parameter at both field orientations. In Fig.~\ref{Fig::Tc_vs_M}a) green and blue dots show these two critical temperatures. When the nematicity axis is perpendicular to the field, the critical temperature is insensitive to the magnetization magnitude. When the magnetization is applied along the nematicity axis, it suppresses critical temperature. Dependence of two order parameters $\Delta_x$ and $\Delta_y$ on the field $m_x$ in the insert in Fig.~\ref{Fig::Tc_vs_M}a) shows that critical temperature $\Delta_y$ is independent of $m_x$ while critical temperature of $\Delta_x$ decreases with of the increase of $m_x$ . 

The state with the higher critical temperature is the most energetically favorable near the critical temperature. So, without warping or pinning fields, the in-plane magnetization selects the orthogonal direction of the nematicity. In means that for $m_x$ magnetization, the most favorable order parameter is $\Delta_{y}$ while for $m_y$ most favorable is $\Delta_{x}$.



\section{Bulk Fermi surface evolution by the magnetization}
In the previous section, we have shown that E$_u$ superconductivity is robust with respect to magnetization with the magnitude $\sim 10$~-~$100\: T_c$. Now, we investigate how such a magnetizztion influences the bulk spectrum of the nematic superconductor. We show that strong magnetization changes the topology of the Fermi surface. 

For convenience, we focus on the orientation $\Delta_y$ of the order parameter that corresponds to the nematicity $\alpha=\pi/2$ since it has the lowest free energy among possible order parameters~(\ref{Eq::Delta}) for zero magnetic field~\cite{Fu2014,Akzyanov2020}. Also, we focus on the plane $k_z=0$ since the most crucial changes in the spectrum occur there.

\subsection{Spectrum without warping}
\begin{figure}[h]
    \center{\includegraphics[width=1\linewidth]{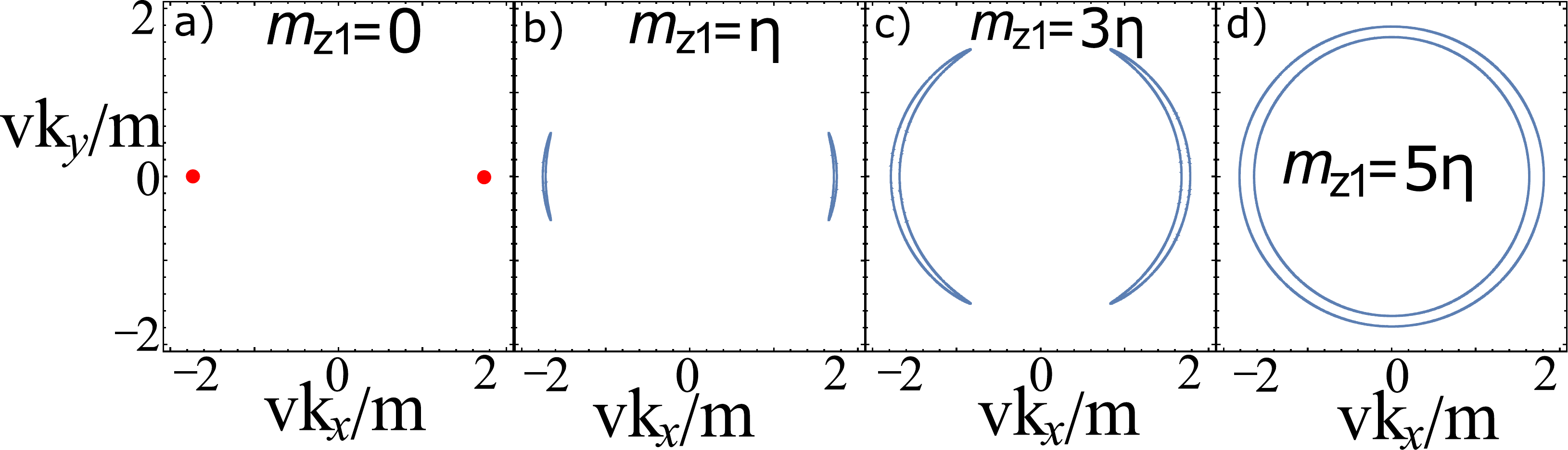}}
    \caption{ Fermi surface of the of nematic superconductor $E_{\mathbf{k}}=0$ with $\Delta_y$ orientation in dimensionless momentum cooridantes $(vk_x/m,vk_y/m)$. The figure is plotted without the hexagonal warping ($\lambda=0$). The magnitude of the order parameter is $\eta$. a) Without magnetization Fermi surface consists of two nodal points. b) Small magnetization $m_z=\eta$ transforms each nodal point to a small Fermi surface. b) Larger magnetization values $m_z=3\eta$ increase the area of the Fermi surface. c) At $m_z=5\eta$, Fermi surfaces merge into two split circles from each nodal point.}
    \label{Fig::L=0}
\end{figure}

We start from the model without the hexagonal warping (i.e., $\lambda=0$). Without the hexagonal warping, the system has a $C_{\infty}$ rotational symmetry: simultaneous rotation of the order parameter, and the magnetization does not change physical properties. In the zero Zeeman field, the spectrum has two nodal points (Fig.~\ref{Fig::L=0}a). Turning on $m_{z}$ leads to a spectrum transformation. Two nodal points transform into a pair of closed nodal lines (Fig.~\ref{Fig::L=0}b). Further, an increase in the Zeeman field makes these nodal lines bigger (Fig~\ref{Fig::L=0}). Finally, they become two nested closed circles (Fig.~\ref{Fig::L=0}d). 
\subsection{Spectrum with warping}
Now we include the hexagonal warping in our model. We have the full gap for the order parameter $\Delta_y$ and in zero magnetization. In contrast with the model without hexagonal warping, magnetization can close the gap only in several nodal points. We consider the model with hexagonal warping and investigate the spectrum for different orientations of the Zeeman field.

\begin{figure}[h]
    \center{\includegraphics[width=1\linewidth]{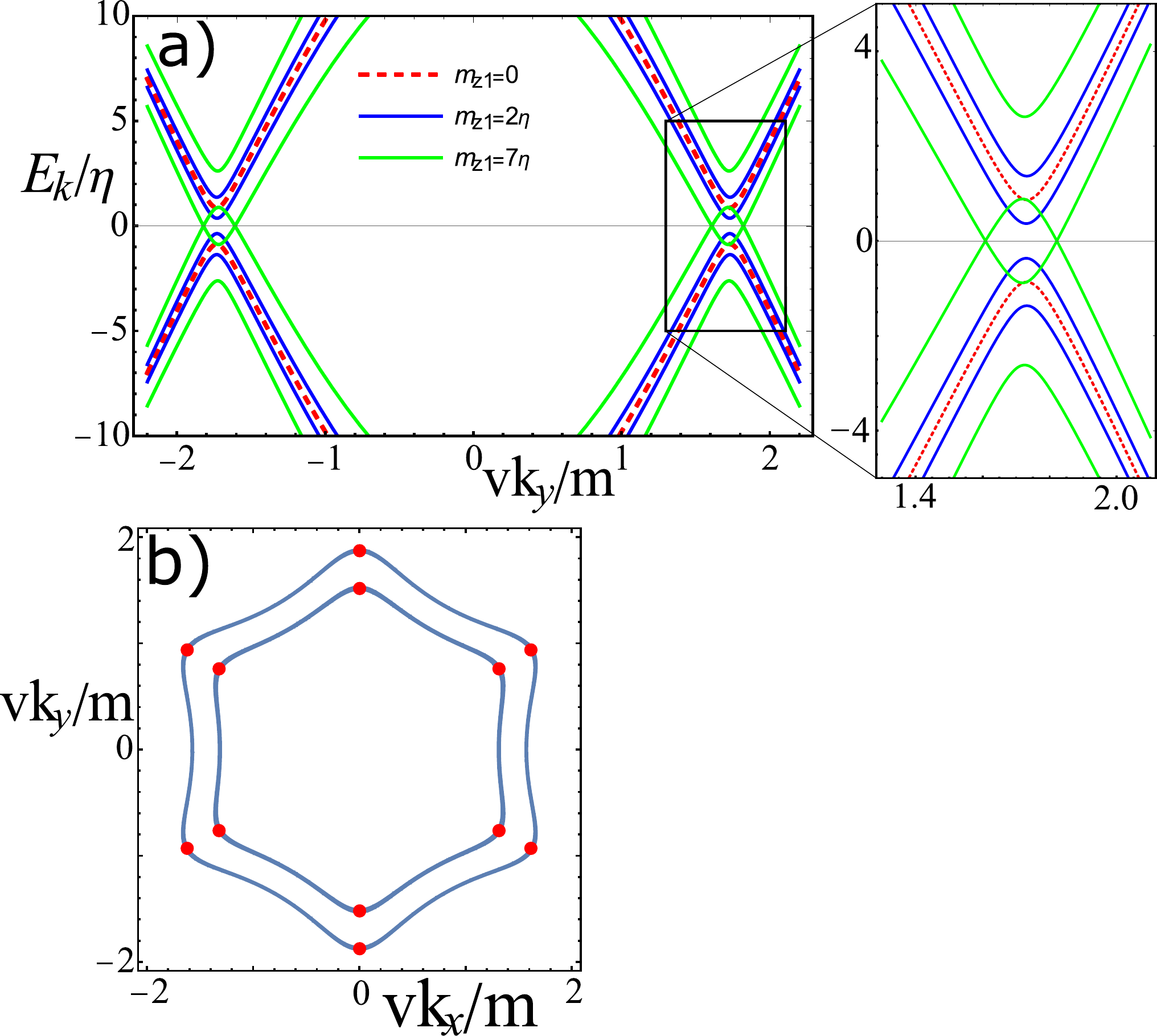}}
    \caption{Panel a) : Dimensionless spectrum of the nematic superconductor $E_{\mathbf{k}}/\eta$ vs dimensionless momentum $vk_y/m$. Order parameter of magnitude $\eta$ has $\alpha=\pi/2$ ($\Delta_y$) orientation. We set $\mu=2m$ and $\lambda m^2/v^3=0.3$. Dash red line gives gaped spectrum without Zeeman field $m_{z}=0$. The blue line corresponds to a small Zeeman field of magnitude $m_{z}=2\eta$ that is too small to close the gap. The green line gives the system in the presence of a strong Zeeman field $m_{z}=7\eta$ that closes the gap at 12 nodal points. Panel b): Fermi surface for the normal state in plane $k_z=0$ with nonzero $m_{z}>0$. Red points indicate nodal points in the presence of superconductivity.}
    \label{Fig::mz}
\end{figure}

We set $m_{x}=m_{y}=0$ in Eq.~(\ref{Eq::Hm}) and consider only $m_z$ for now. Low-energy band of $E_{\mathbf{k}}$ splits into two bands with energies $E_{\mathbf{k}}\pm\delta\epsilon(\mathbf{k})$. Low magnetization lifts the degeneracy keeping the gap open. See the blue curve in Fig.~\ref{Fig::mz}a). A strong field makes splitting between bands higher. Finally, electron and hole bands cross at the Fermi level, and the gap closes at 12 nodal points (Fig.~\ref{Fig::mz}b).    

\begin{figure}[h]
    \center{\includegraphics[width=1\linewidth]{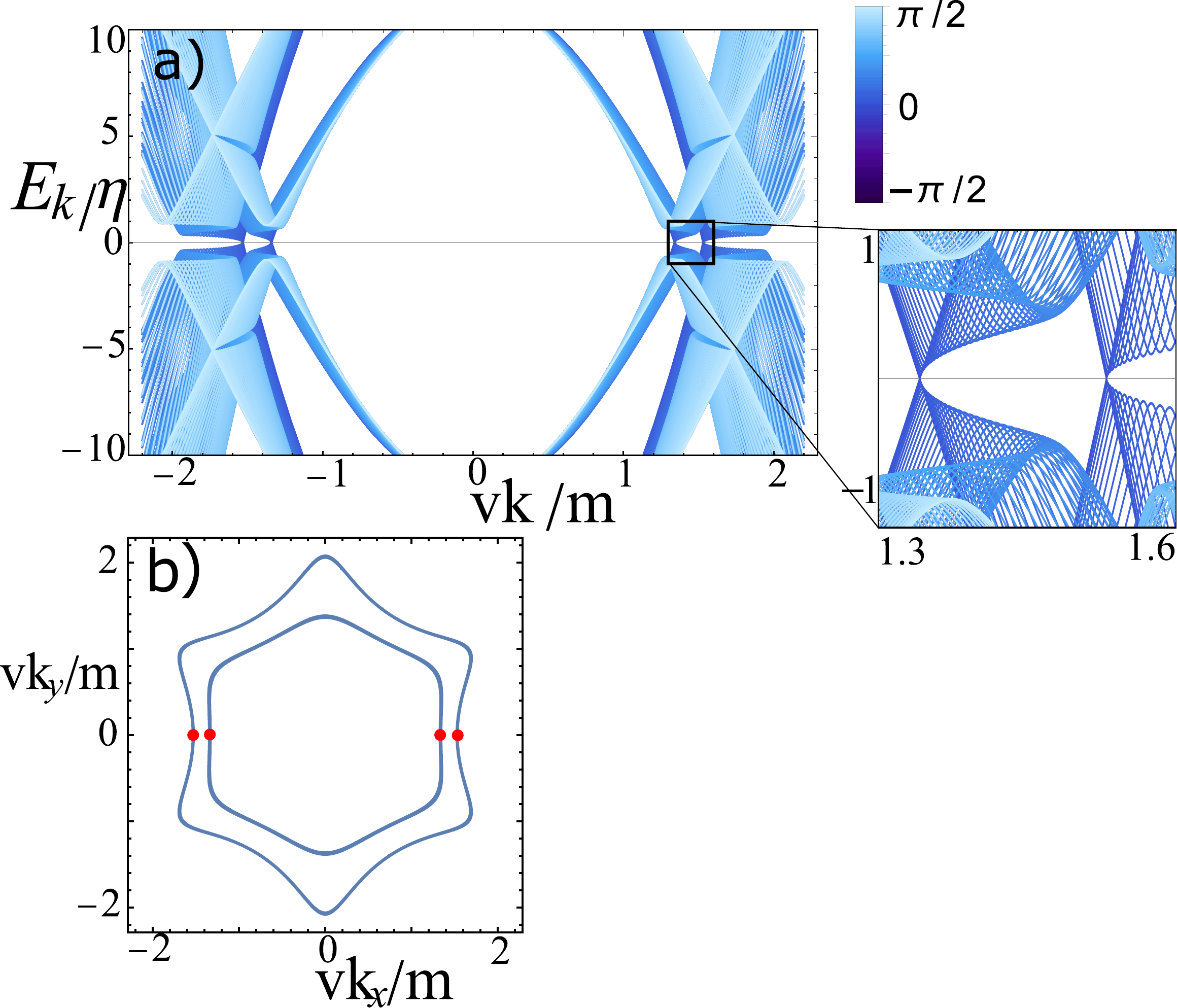}}
    \caption{a): Dimensionless spectrum of nematic superconductor $E_{\mathbf{k}}/\eta$ vs dimensionless momentum $vk/m$. Order parameter of magnitude $\eta$ has $\Delta_y$ orientation. Different colors correspond to different orientations of the spectrum's cut. The spectrum is plotted with the following parameters: $m_x=10\eta,\,\,\mu=2m,\,\,\lambda m^2/v^3=0.3$. b): Fermi surface in plane $k_z=0$ with nonzero $m_{x}$ and without superconductivity. Red points indicate nodes in the presence of superconductivity.}
    \label{Fig::mx}
\end{figure}

We apply magnetization $m_x$ along the $Ox$ axis. We show spectrum along the set of vectors $\mathbf{k}$ in Fig.~\ref{Fig::mx}a. Each vector $\mathbf{k}$ is placed in plane $(k_x;k_y)$ and has angle $\beta\in[-\pi/2;\pi/2]$ with $k_x$. When we apply a strong enough field, the gap closes at $\phi=0$ at 4 different nodal points. We indicate nodes by red points at the Fermi surface of the normal phase in Fig.~\ref{Fig::mx}b. 

\begin{figure}[h!]
    \center{\includegraphics[width=1\linewidth]{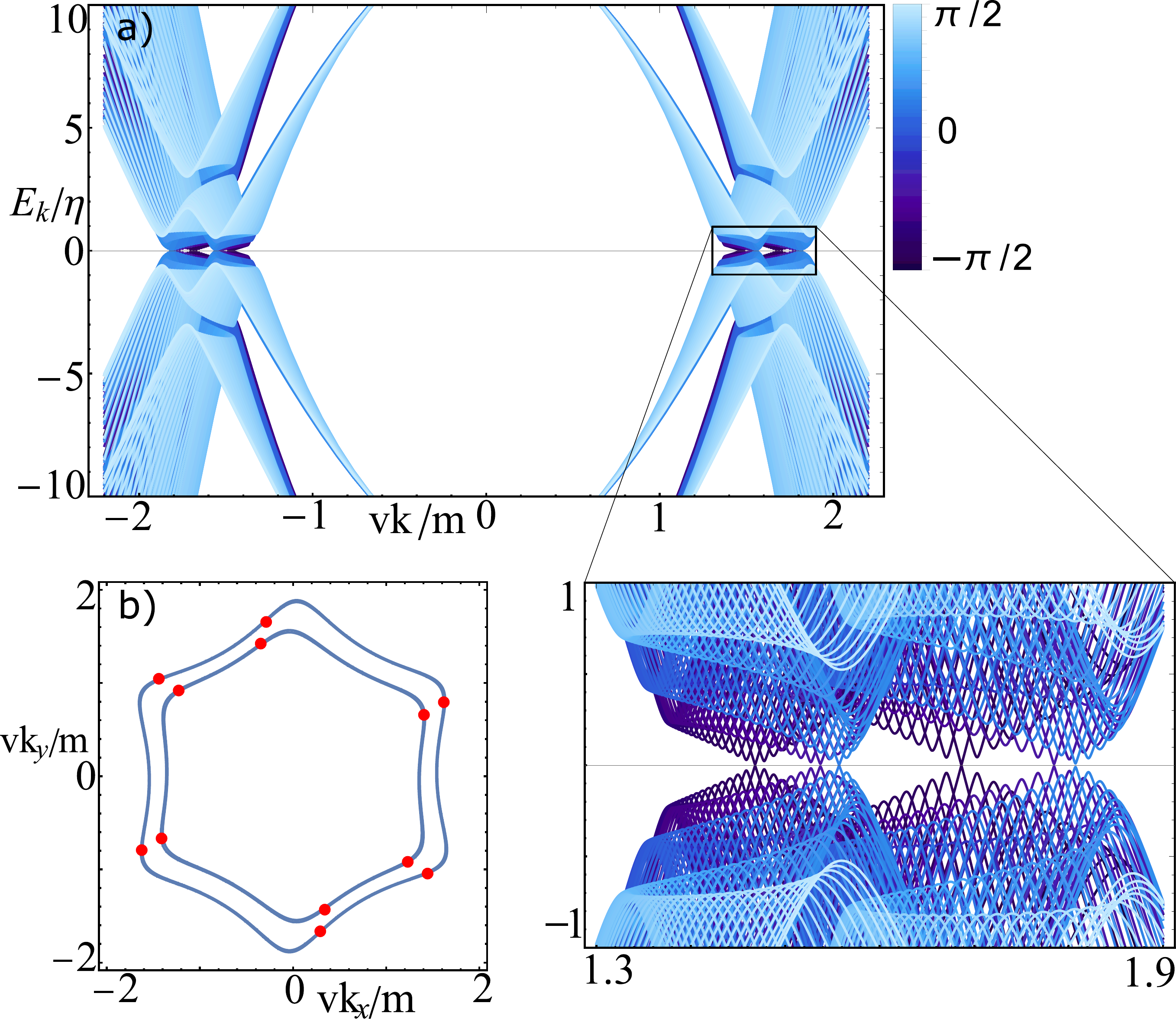}}
    \caption{Panel a): Dimensionless spectrum of nematic superconductor $E_{\mathbf{k}}/\eta$ vs dimensionless momentum $vk/m$. The momentum stays in plane $Oxy$. Different colors give the angle between $\mathbf{k}$ and $k_x$. We set $m_{x}=3\eta,\,\,m_{z}=7\eta,\,\,\mu=2m,\,\,\lambda m^2/v^3=0.3$.  The order parameter of magnitude $\eta$ has $\Delta_y$ orientation.. Insert shows that the gap closes along three $\mathbf{k}$ orientations. Panel b):  Fermi surface in plane $k_z=0$ with nonzero $m_{x}$ and $m_{z}$ and without superconductivity. Red points indicate nodes in the presence of superconductivity.}
\label{Fig::mxmz}
\end{figure}

Now we turn on the Zeeman field in the $Oxz$ plane. We again focus on the $(k_x;k_y)$ plane in the momentum space. When $m_{z}$ is relatively high, we have 12 nodal points (Fig.~\ref{Fig::mxmz}a). The system has the same topology as in the case of pure $m_{z}$ field (Fig.~\ref{Fig::mz}b). Due to the presence of $m_{x}$, these nodal points are shifted from symmetrical positions in the vertices of the Fermi surface (Fig.~\ref{Fig::mxmz}b).

When $m_{z}\lesssim m_{x}$ 8 of 12 nodal points gaped, and the topology of the Fermi surface becomes equal to the case with $m_{x}$ only with 4 nodal points (Fig.~\ref{Fig::mx}a). 

\begin{figure}[h!]
    \center{\includegraphics[width=1\linewidth]{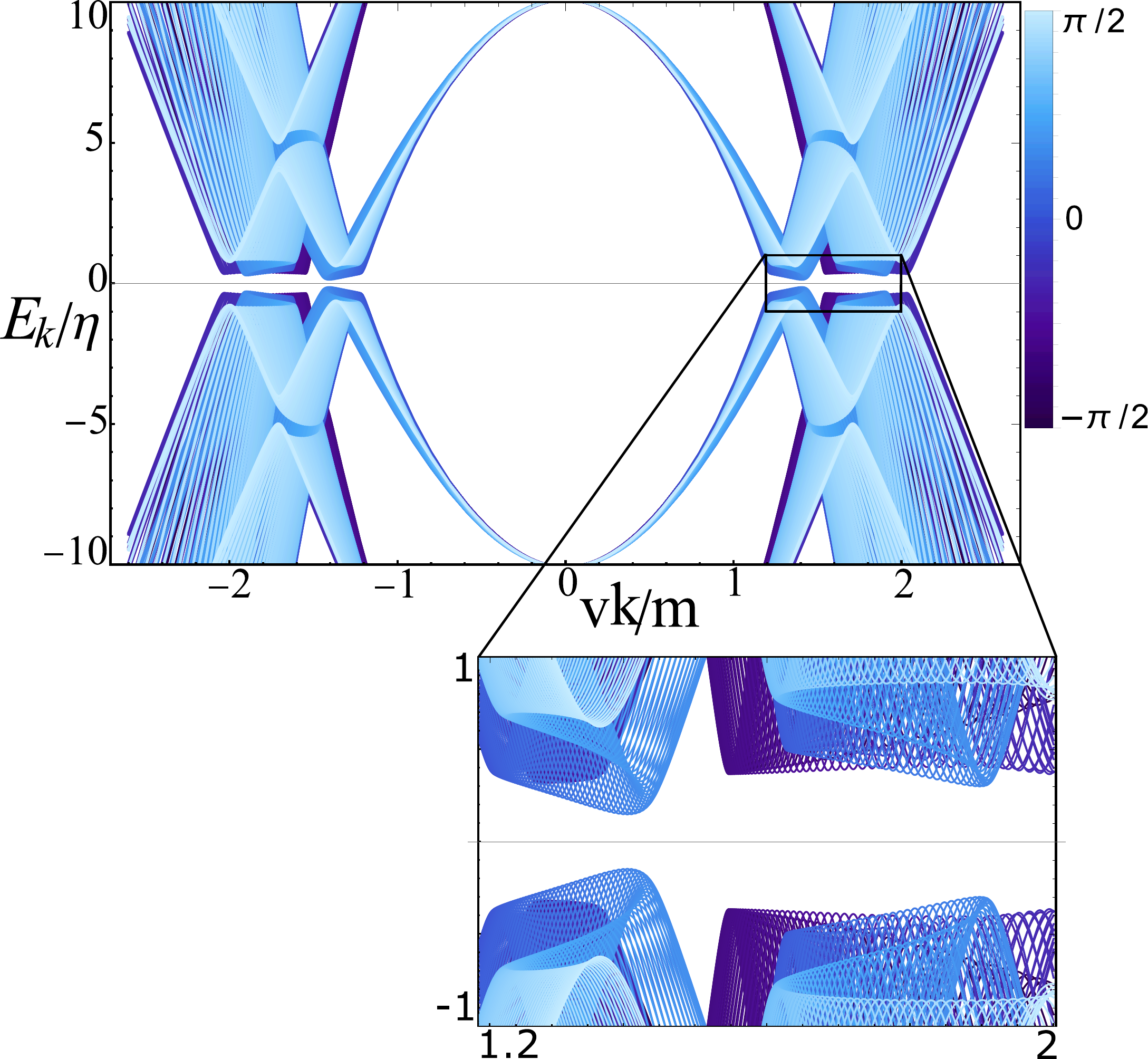}}
    \caption{Panel a): Dimensionless spectrum of nematic superconductor $E_{\mathbf{k}}/\eta$ vs dimensionless momentum $vk/m$. The momentum stays in plane $Oxy$. Different colors give the angle between $\mathbf{k}$ and $k_x$. We set $m_x=7\eta,\,\,m_{y}=10\eta,\,\,m_{z}=10\eta\mu=2m,\,\,\lambda m^2/v^3=0.3$. The order parameter of magnitude $\eta$ has $\Delta_y$ orientation. The full gap is opened. }
\label{Fig::mxmymz}
\end{figure}

 Finite magnetization along $Oy$ direction $m_y$ opens the full gap in the spectrum without any nodal points (Fig.~\ref{Fig::mxmymz}). 
\section{Surface states}
\begin{figure}[h]
    \center{\includegraphics[width=1\linewidth]{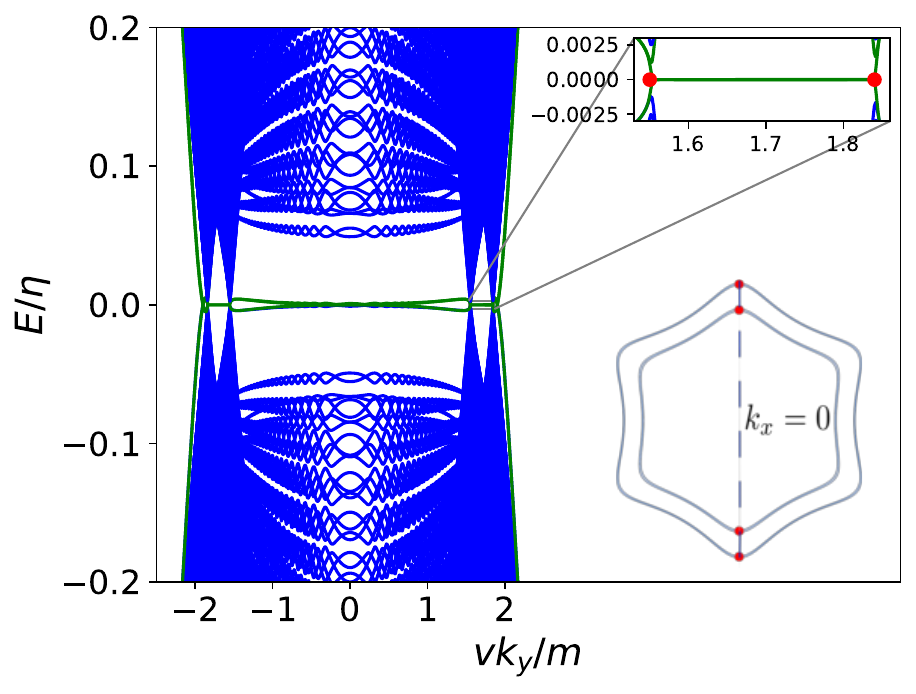}}
    \caption{Dimensionless spectrum of nematic superconductor $E/\eta$ vs dimensionless momentum $vk_x/m$ along the $x$ axis.Order parameter of magnitude $\eta$ has an orientation $\alpha=\pi/2$ ($\Delta_y$). Red dots correspond to nodal points, blue lines are bulk states, and green are surface states. Flat surface states connect nodal points. The spectrum is plotted with the following parameters: $m_x=m_y=0,\,\,m_z=3\eta,\,\,\mu=2m,\,\,\lambda m^2/v^3=0.3$.}
    \label{fig:nodal_kx}
\end{figure}

\begin{figure}[h]
    \center{\includegraphics[width=1\linewidth]{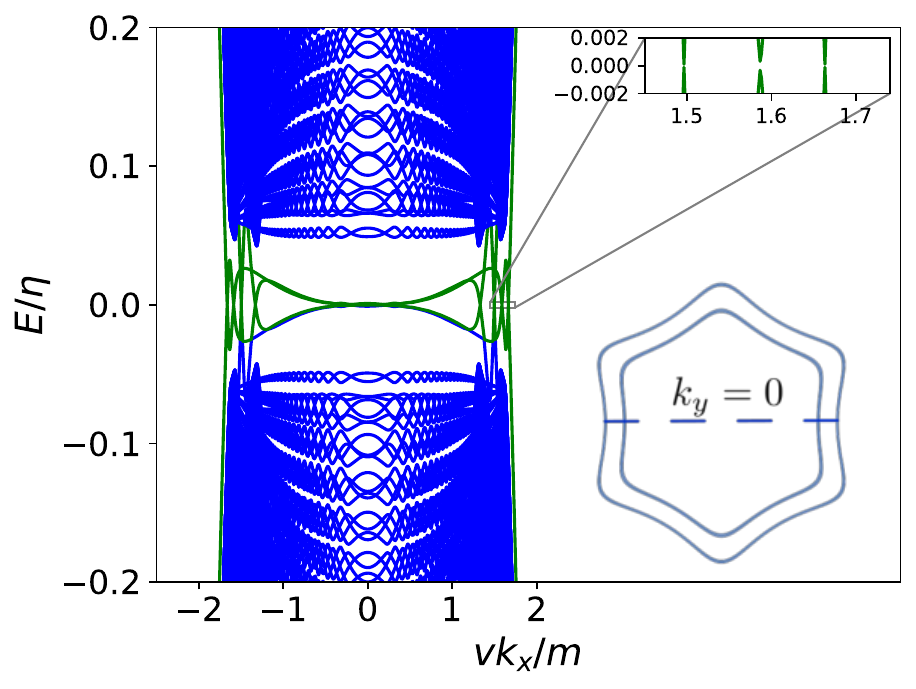}}
    \caption{Dimensionless spectrum of nematic superconductor $E/\eta$ vs dimensionless momentum $vk_y/m$. Order parameter of magnitude $\eta$ has an orientation $\alpha=\pi/2$ ($\Delta_y$). Blue lines are bulk states, and green lines are surface states. There is no flat Majorana nodal line since the cut does not include bulk nodal points. The spectrum is plotted with the following parameters: $m_x=m_y=0,\,\,m_z=3\eta,\,\,\mu=2m,\,\,\lambda m^2/v^3=0.3$.}
    \label{fig:nodal_ky}
\end{figure}

In this section, we briefly analyze the properties of the surface states on the nematic superconductor with the magnetization along the z-axis. We rewrite our Hamiltonian in a tight-binding approximation with the additional quadratic terms that renormalize chemical potential $\mu \rightarrow \mu+ C_i k_i^2$ and single electron gap $m \rightarrow m+ B_i k_i^2$, $i=x,y,z$, see Ref.~\cite{Liu2010} for details on the values of the parameters. We calculate the spectrum for $n=200$ layers of the doped topological insulator stacked along the $Oz$ axis. In this case, we can catch surface states along with the axis of the bulk states of the superconductor. The spectrum of the nematic superconductor along $x$ and $y$ axes is shown in Figs.~\ref{fig:nodal_kx} and ~\ref{fig:nodal_ky} respectively. 

In Fig.~\ref{fig:nodal_kx}, we show the cut that takes nodal points. We see that between nodal points, the spectrum is similar to the spectrum of the surface states without the magnetization\cite{Hao2017,Khokhlov2022} . A flat Majorana nodal line connects nodal points. This line is very flat up to numerical error that arises in the calculations. In Fig.~\ref{fig:nodal_ky}, we can see that no such flat lines occur at the slice that does not catch bulk nodal points. Also, such flat lines that connect nodal points occur at all orientations of the magnetic fields where such nodal points occur.
\section{Discussion}
We investigated critical temperature of spin triplet E$_u$ superconductivity with vector order parameter, that appears in doped Bi$_2$Se$_3$. We find, this superconductivity exist for large values of the magnetization of about $10$-$100$ $T_c$ which is typical for the spin-triplet superconductivity. 

One of the main results that we can tune the direction of the nematic order parameter by the external magnetization, see Fig.~\ref{Fig::Tc_vs_M}. In the recent experiment~\cite{Yokoyama2023} it was shown that for the low values of the in-plane magnetic field the direction of the nematicity is fixed which is consistent with the previous studies~\cite{Kuntsevich2019}. However, at large enough values of the magnetic field the nematicity orientation is no longer pinned and starts following the direction of the magnetic field. These results are consistent with our predictions that in-plane magnetic field chooses the direction of the order parameter, see Fig.~\ref{Fig::Tc_vs_M}.

We have shown that magnetization can open nodal points in the bulk spectra. Such opening occurs either for the magnetization along $0z$ direction, see Fig.~\ref{Fig::mz}, or magnetization perpeniducular to the nematicity direction, see Fig.~\ref{Fig::mx}, or both, see Fig.\ref{Fig::mxmz}. However, magnetization parallel to the direction of the order parameter lifts the nodal points, see Fig.\ref{Fig::mxmymz}. These features can be experimentally measured. In case of the full gap, the variation of the London penetration length as a function of a temperature should be exponential $\delta \lambda_L/\lambda_L \propto e^{-\textrm{gap}/T}$. Nodal points change law from the exponential to the quadratic $\delta \lambda_L/\lambda_L \propto T^2$~\cite{Smylie2022}. So, different asymptotics for the London penetration length for the different orientations of the magnetic fields would be an experimental justification of our theory.

In conclusion, we have shown that nematic superconductivity in doped topological insulators is robust against the magnetization. The direction of the nematic order parameter can be tuned by the direction of the in-plane magnetization. Large values of the out-of-plane magnetization favor chiral superconducting state. In case of the pinned nematic state, the magnetization can open nodal points in the bulk spectrum. Such nodal points are split by the Zeeman field. Splitted nodal points are connected through the flat nodal line Majorana surface states. Magnetization parallel to the direction of the order parameter lifts the nodal points.


\section*{Acknowledgment}
Authors acknowledge support by the Russian Science Foundation under Grant No 20-72-00030. AVK thanks the partial support from the Foundation for the Advancement of Theoretical Physics and Mathematics “BASIS”. 

\bibliography{nematic}
\end{document}